\documentclass[sigconf]{acmart}

\usepackage{hyperref}
\usepackage{booktabs}
\usepackage{multirow}
\usepackage{makecell}
\usepackage{rotating}
\usepackage{amsthm}
\usepackage{graphicx}
\usepackage{subcaption}
\usepackage{soul}
\usepackage{balance}
\usepackage{xcolor}
\usepackage{lipsum}

\newsavebox{\stripboxbox}
\newenvironment{stripbox}[2][gray!10]{%
  \def\stripbg{#1}%
  \def\stripvspace{#2}%
  \par\vspace{\stripvspace}\noindent %
  \begingroup
  \setlength{\fboxsep}{0pt}%
  \begin{lrbox}{\stripboxbox}%
    \begin{minipage}{\dimexpr\linewidth-4pt\relax}%
      \hspace*{2pt}%
      \begin{minipage}{\dimexpr\linewidth-5pt\relax}%
}{%
      \end{minipage}%
      \hspace*{2pt}%
    \end{minipage}%
  \end{lrbox}%
  \colorbox{\stripbg}{%
    \parbox{\linewidth}{%
      \noindent\rule{\linewidth}{1pt}\par %
      \vspace{2pt}%
      \noindent\hspace*{2pt}\usebox{\stripboxbox}\hspace*{2pt}\par
      \vspace{2pt}%
      \noindent\rule{\linewidth}{1pt}%
    }%
  }%
  \endgroup
  \par\vspace{\stripvspace}%
}

\theoremstyle{definition}   
\newtheorem{definition}{Definition}

\newboolean{showcomments}
\setboolean{showcomments}{true}
\ifthenelse{\boolean{showcomments}}
  {\newcommand{\nb}[2]{
  \fbox{\bfseries\sffamily\scriptsize#1}
     {\sf\small$\blacktriangleright$\textit{\textcolor{red}{#2}}$\blacktriangleleft$}
   }
  }
  {\newcommand{\nb}[2]{}
   
  }

\newcommand\changed[1]{{\color{black}#1}} 

\AtBeginDocument{%
  }

\copyrightyear{2026}
\acmYear{2026}
\setcopyright{cc}
\setcctype{by}
\acmConference[ICSE '26]{2026 IEEE/ACM 48th International Conference on Software Engineering}{April 12--18, 2026}{Rio de Janeiro, Brazil}
\acmBooktitle{2026 IEEE/ACM 48th International Conference on Software Engineering (ICSE '26), April 12--18, 2026, Rio de Janeiro, Brazil}
\acmPrice{}
\acmDOI{10.1145/3744916.3773148}
\acmISBN{979-8-4007-2025-3/26/04}

\author{Jinhan Kim}
\orcid{0000-0002-0140-7908}
\affiliation{%
  \institution{Università della Svizzera italiana}
  \city{Lugano}
  \country{Switzerland}}
\email{jinhan.kim@usi.ch}

\author{Nargiz Humbatova}
\orcid{0000-0002-3037-8368}
\affiliation{%
  \institution{Università della Svizzera italiana}
  \city{Lugano}
  \country{Switzerland}}
\email{nargiz.humbatova@usi.ch}

\author{Gunel Jahangirova}
\orcid{0000-0002-1423-1083}
\affiliation{%
  \institution{King's College London}
  \city{London}
  \country{UK}}
\email{gunel.jahangirova@kcl.ac.uk}

\author{Shin Yoo}
\orcid{0000-0002-0836-6993}
\affiliation{%
  \institution{KAIST}
  \city{Daejeon}
  \country{Republic of Korea}}
\email{shin.yoo@kaist.ac.kr}

\author{Paolo Tonella}
\orcid{0000-0003-3088-0339}
\affiliation{%
  \institution{Università della Svizzera italiana}
  \city{Lugano}
  \country{Switzerland}}
\email{paolo.tonella@usi.ch}

\begin{document}

\title[A Critical Review and Implications on DNN Coverage Testing]{Revisiting ``Revisiting Neuron Coverage for DNN Testing: A Layer-Wise and Distribution-Aware Criterion'': A Critical Review and Implications on DNN Coverage Testing}

\begin{abstract}
  We present a critical review of Neural Coverage (NLC), a state-of-the-art DNN coverage criterion by Yuan et al. at ICSE 2023. While NLC proposes to satisfy eight design requirements and demonstrates strong empirical performance, we question some of their theoretical and empirical assumptions. We observe that NLC deviates from core principles of coverage criteria, such as monotonicity and test suite order independence, and could more fully account for key properties of the covariance matrix. Additionally, we note threats to the validity of the empirical study, related to the ground truth ordering of test suites. Through our empirical validation, we substantiate our claims and propose improvements for future DNN coverage metrics. Finally, we conclude by discussing the implications of these insights.
\end{abstract}



\maketitle

\section{Introduction}
\label{sec:introduction}

Testing of Deep Neural Networks (DNNs) has become a crucial aspect of their integration into various applications, sparking a surge in the development of coverage criteria. These criteria draw inspiration from traditional software testing methodologies, aiming to ensure the reliability and robustness of DNNs. A significant milestone in this field was the introduction of Neuron Coverage (NC) by DeepXplore in SOSP 2017~\cite{pei2017deepxplore}, marking the first coverage criterion specifically tailored for DNNs. This drew considerable interest within the Software Engineering (SE) community, leading to the creation of various coverage criteria. Such criteria typically analyse the \textit{activation space of inner neurons} and quantify the extent to which this space has been covered~\cite{ma2018deepgauge, kim2019guiding}.

At ICSE 2023, a new DNN coverage criterion called Neural Coverage (NLC) was introduced~\cite{nlcpaper}. NLC distinguishes itself from prior coverage criteria by proposing and satisfying eight design requirements, which a DNN coverage criterion is expected to satisfy. To date, NLC is the only criterion that fulfils all these requirements. As a hyperparameter-free metric, NLC characterises the correlation between neurons' continuous outputs. Empirical studies conducted by the NLC proponents have demonstrated that NLC outperforms other coverage criteria, underscoring the strength of its novel design~\cite{nlcpaper}. The implementation of NLC, along with that of existing coverage criteria, and all empirical results are publicly available, which has gained significant traction, as evidenced by its 250+ GitHub stars.\footnote{\url{https://github.com/Yuanyuan-Yuan/NeuraL-Coverage}, accessed on February 11, 2025.}

Despite the proponents' claims and the empirical evidence, we question some of the theoretical and empirical assumptions behind the work by Yuan et al.~\cite{nlcpaper}.\footnote{Throughout the paper, we refer to Yuan et al.~\cite{nlcpaper} simply as Yuan et al. (without repeated citations).} %
As suggested by Monperrus~\cite{monperrus2014critical}, we believe that respectful critique and scholarly debate are essential to the scientific process. While such discussions often occur within technical contributions (e.g., in related work sections), they are frequently superficial and biased toward the new approach. Explicit discussion of limitations and open issues in a fully dedicated paper, on the other hand, can provide a clearer understanding of the area, pointing to better ways to advance the discipline. %

Our critical review is driven by two primary motivations: (1) concerns regarding the theoretical foundations of NLC, and (2) concerns about its empirical validation. First, we articulate our conceptual concerns by analysing NLC's design in light of the eight proposed requirements. \changed{Specifically, we highlight three major shortcomings: the lack of \textit{monotonicity} (i.e., adding more tests should not decrease coverage), the lack of \textit{order independence} (i.e., the coverage value should remain unchanged under permutations of the test suite), and the absence of a \textit{bounded coverage measure} (i.e., the coverage value should have a known upper bound to serve as a stopping criterion). These are essential properties for any reliable DNN coverage criterion.} In detail, we explore why NLC fails to serve as a sound coverage criterion, discuss the trade-offs involved in satisfying the considered requirements, address potential misconceptions about NLC's relationship with the covariance matrix, and propose improvements for future work. Second, we identify flaws in the empirical study design and offer recommendations for the rigorous evaluation of DNN coverage criteria.

One major concern descends from their requirement R1, stating: \textit{`DNN coverage criterion should precisely describe the continuous output of a neuron.'}. The authors argue that most existing criteria fail to satisfy R1 because they rely on thresholding, which artificially divides the neuron output space and fails to capture its true shape. \changed{While this claim is valid, we contend that by working without any thresholding, NLC has turned into a non-monotonic, unbounded coverage criterion, whose value may decrease as new tests are added to a test suite and without any clearly defined maximum value. Correspondingly, it cannot function as a reliable coverage criterion.} The authors partially recognised this issue and proposed a fix, which involves updating coverage only when it increases. However, this solution introduces a dependency on the order of test case evaluation, as the same test case may or may not be taken into account by NLC, depending on the sequence of evaluation. We argue that monotonicity and order independence should be intrinsic properties of the coverage criterion's design. Furthermore, we empirically validate our argument regarding order dependency, demonstrating that it is a practically significant issue. Our experiments reveal that NLC coverage can decrease by up to 10.42\% depending on the evaluation order.

In addition to our theoretical review of NLC, we critically analyse the empirical study design, which significantly deviates from prior work~\cite{pei2017deepxplore,kim2019guiding,ji2023cc,ma2018deepgauge}. Their approach establishes a ground truth ordering of test suites, for instance, asserting that test suite $T_1$ must have higher coverage than another suite $T_2$ based on their posit that $T_1$ exhibits greater diversity than $T_2$. This is used to evaluate and compare coverage criteria in the absence of definitive ground truth in the literature. We highlight substantial threats to the validity of their study. This is primarily because the assumed ground truth does not always hold, as coverage cannot be equated to diversity,
rendering their evaluation unreliable.

At the end of the paper, we revisit the critiques outlined earlier, provide additional insights, and suggest potential solutions to the issues identified in Yuan et al.'s work, concluding with key takeaways.
In summary, our contributions are:
\begin{itemize}
    \item An analysis of how NLC misses important properties of coverage criteria. We examine the eight proposed requirements, the preservation of covariance matrix properties, and the proposed layer-wise aggregation.
    \item Identification of critical threats to the validity in their empirical study design, particularly in establishing a ground truth ordering of test suites.
    \item Empirical validation of our arguments to substantiate our claims.
    \item A comprehensive discussion, guiding an improved design and evaluation of future DNN coverage criteria.
\end{itemize}

\section{Background}
\label{sec:background}

In this section, we provide an overview of DNN coverage testing and delve into the work by Yuan et al., which introduced NLC, a novel coverage criterion designed to address the limitations of existing metrics.

\subsection{DNN Coverage Testing}
\label{sec:dnn_coverage_testing}

DNN coverage testing evaluates the extent to which a DNN's internal behaviour is exercised during testing. Various criteria have been proposed, each targeting different aspects of DNN behaviour.

One class of criteria focuses on individual neurons. For instance, \textit{Neuron Coverage} (NC)~\cite{pei2017deepxplore} measures the proportion of neurons activated beyond a predefined threshold. More advanced methods, such as \textit{k-Multisection Neuron Coverage} (KMNC)~\cite{ma2018deepgauge}, divide the output range of each neuron into sections and track coverage based on which sections are activated. Additional methods like \textit{Neuron Boundary Coverage} (NBC) and \textit{Strong Neuron Activation Coverage} (SNAC) focus on neurons with outputs outside expected ranges or exceeding upper bounds, respectively, to identify extreme or anomalous behaviours~\cite{ma2018deepgauge}. Another set of criteria examines groups of neurons within layers. For example, \textit{Top-k Neuron Coverage} (TKNC)~\cite{ma2018deepgauge} identifies the top $k$ neurons with the highest outputs in each layer. \textit{Surprise Coverage} (SC)~\cite{kim2019guiding, kim2023evaluating} compares neuron traces of test inputs to those from training data, using distance metrics such as Mahalanobis distance to quantify differences (MDSC). Variants like LSC and DSC employ alternative distance measures. \textit{Causal Coverage} (CC)~\cite{ji2023cc} captures causal relationships between neurons through statistical independence tests. \changed{LidSA~\cite{guo2024neuron} improves SC by estimating the Local Intrinsic Dimensionality (LID)~\cite{houle2017local} of activation traces from important neurons, identified via contribution propagation. Two associated criteria, \textit{LidSA Surprise Coverage} (LDSC) and \textit{Top-N Surprise Coverage} (TNSC), assess how well a test suite covers the range of LidSA values.
\textit{Input Domain Coverage} (IDC)~\cite{dola2023input} introduces a black-box approach to test adequacy by leveraging a Variational Autoencoder (VAE) to map inputs into a latent feature space. Combinatorial Interaction Testing (CIT) is then applied in this space to measure how well test inputs cover combinations of latent features. IDC shows a strong correlation with fault detection and complements white-box criteria by capturing input diversity.} \textit{Neuron Path Coverage} (NPC)~\cite{xie2022npc} represents the internal decision structure of the DNN on given inputs, analogous to control flow in traditional software.
Our reference work, NLC~\cite{nlcpaper}, stands out as a unique criterion by addressing specific design requirements for DNN testing. It is the only criterion that fulfils all these requirements, taking into account the correlation, distribution, and density of each layer's outputs in a DNN, while also supporting various optimisations. We provide a detailed explanation of NLC in the subsequent section.

\subsection{NLC by Yuan et al.~\cite{nlcpaper}} 
In this section, we introduce the requirements proposed by Yuan et al. and discuss how NLC is designed to satisfy these requirements.

\subsubsection{Suggested Requirements}
To address the limitations of existing neuron coverage criteria, Yuan et al. proposed eight design requirements for DNN coverage criteria. %
Below, we introduce and explain each requirement in detail.

\textbf{R1: \textit{DNN coverage should precisely describe the continuous output of a neuron.}}
DNNs process inputs through continuous, non-linear transformations, and the outputs of neurons are continuous values. Existing coverage criteria often discretise these outputs (e.g., by applying thresholds) to simplify analysis. However, this approach is reductive, as it artificially partitions the output space and disregards the continuous nature of neuron activations. R1 emphasises that a DNN coverage criterion should accurately describe the continuous output of a neuron, ensuring that the metric reflects the true behaviour of the DNN.

\textbf{R2: \textit{It is preferable to measure DNN coverage by characterising correlations of neurons.}}
Neurons within a DNN layer are interconnected, meaning their outputs are correlated and collectively contribute to processing input features. Existing criteria often treat neurons as independent units, which is inaccurate because the behaviour of one neuron can influence others. R2 highlights the importance of measuring DNN coverage by characterising the correlations between neurons. %

\textbf{R3: \textit{DNN coverage should analyse how outputs of neurons in a layer are distributed.}}
DNNs operate by approximating complex distributions through their hierarchical layers, with each layer responsible for extracting and refining features at varying levels of abstraction. R3 posits that a DNN coverage criterion should analyse the distribution of neuron outputs within each layer. 

\textbf{R4: \textit{DNN coverage should consider density of neuron output distributions.}}
The outputs of neurons often cluster within the output space, with regions of varying density. High-density regions typically correspond to common or expected behaviours, while low-density regions may signify rare or anomalous patterns. R4 mandates that a DNN coverage criterion consider the density of neuron output distributions. %

\textbf{R5: \textit{DNN coverage should support incorporating prior knowledge extracted from training data.}}
Training data reflects how a DNN learns to approximate distributions, and this knowledge can be leveraged to refine coverage metrics. R5 emphasises that a DNN coverage criterion should support the incorporation of prior knowledge extracted from training data. %

\textbf{R6: \textit{DNN coverage should support matrix-form computation to be optimisable by modern DL frameworks and hardware.}}
Modern deep learning frameworks, such as PyTorch and TensorFlow, along with hardware accelerators like GPUs, are optimised for matrix operations. R6 stipulates that a DNN coverage criterion should support matrix-form computation to capitalise on these optimisations for large-scale DNNs.

\textbf{R7: \textit{DNN coverage should feature efficient incremental update.}}
In practice, test inputs are often generated and evaluated incrementally, such as in fuzz testing or online testing scenarios. R7 requires that a DNN coverage criterion feature efficient incremental updates, without requiring a full recomputation of the metric.

\textbf{R8: \textit{DNN coverage is desirable to be hyperparameter free.}}
Many existing coverage criteria rely on hyperparameters, such as thresholds or bucket sizes, which must be manually configured. R8 advocates for coverage criteria to be hyperparameter-free.

The eight requirements designed by Yuan et al. provide an appealing framework for the design of DNN coverage criteria. By addressing these requirements, NLC emerges as a promising metric that captures the continuous, interdependent, and distributional nature of DNN behaviours. Its ability to incorporate prior knowledge, support efficient computation, and operate without hyperparameters makes it potentially a valuable tool for evaluating and enhancing the quality of DNN test suites.

\subsubsection{Design of DNN Coverage Criterion}
\label{sec:background_design_nlc}

Neural Coverage (NLC) is a layer-wise and distribution-aware metric designed to evaluate the quality of test suites for DNNs. Here, we provide a detailed technical explanation of how NLC is calculated. Neuron outputs are represented as continuous numbers, and NLC directly operates on these outputs in their continuous form without any discretisation. For a single neuron $n$, NLC quantifies the divergence of its output using variance, which measures the spread of the neuron's outputs. The variance $\sigma_n^2$ for neuron $n$ is defined as:

\begin{equation} \label{eq:single_neuron}
    \sigma_n^2 = \mathbb{E}[(o_n - \mathbb{E}[o_n])^2] 
\end{equation}

\noindent where $o_n$ represents the output of neuron $n$, and $\mathbb{E}[o_n]$ is the expected value (mean) of the neuron's output. A higher variance (i.e., higher NLC) indicates that the neuron's outputs are more spread out, suggesting greater activation diversity.

In practice, DNNs typically consist of multiple neurons, and neurons within a layer are interconnected, meaning their outputs are correlated and jointly process input features. To capture this correlation, NLC uses covariance to measure the joint variability of neuron outputs. For two neurons $n_1$ and $n_2$, the covariance $\varsigma_{n_1, n_2}$ is computed as:

\begin{equation}
    \varsigma_{n_1, n_2} = \mathbb{E}[(o_{n_1} - \mathbb{E}[o_{n_1}])(o_{n_2} - \mathbb{E}[o_{n_2}])]
\end{equation}

The covariance matrix $\Sigma$ for a layer with $m$ neurons is then constructed as:

\begin{equation}
    \Sigma = \begin{bmatrix}
    \sigma_{n_1}^2 & \varsigma_{n_1, n_2} & \cdots & \varsigma_{n_1, n_m} \\
    \varsigma_{n_2, n_1} & \sigma_{n_2}^2 & \cdots & \varsigma_{n_2, n_m} \\
    \vdots & \vdots & \ddots & \vdots \\
    \varsigma_{n_m, n_1} & \varsigma_{n_m, n_2} & \cdots & \sigma_{n_m}^2
    \end{bmatrix}
\end{equation}

This covariance matrix not only captures the divergence of individual neurons and their correlations but also encapsulates the shape of the distribution. As a result, NLC leverages the values in the covariance matrix to compute coverage. For a DNN with multiple layers, NLC calculates the coverage for each layer independently and then aggregates the results. The coverage for a single layer $l$ with $m_l$ neurons is computed as:

\begin{equation} \label{eq:nlc_l}
\text{NLC}_l = \frac{1}{m_l \times m_l} \|\Sigma_l\|_1 = \frac{1}{m_l \times m_l} \sum_{j=1}^{m_l} \left( \sum_{i=1}^{m_l} |\varsigma_{n_i, n_j}| \right)
\end{equation}

\noindent where $\Sigma_l$ is the covariance matrix for layer $l$, and $\|\Sigma_l\|_1$ is the L1 norm of the matrix. The overall NLC for the entire DNN is the sum of the coverage values across all layers:

\begin{equation} \label{eq:nlc}
\text{NLC} = \sum_l \text{NLC}_l
\end{equation}

A higher NLC value is expected to indicate a more diverse and comprehensive test suite, as it reflects a greater exploration of the neuron output distributions across all layers of the DNN. %

\section{Critical Review of NLC}
\label{sec:critical_review_nlc}

In this section, we elaborate on our concerns. While our critique focuses on NLC, the discussion aims to offer broader insights into the design of any novel DNN coverage criteria.

\subsection{Does NLC Qualify as a Coverage Criterion?}
\label{sec:is_nlc_valid}

A fundamental property of any test coverage criterion is that coverage should be non-decreasing as new tests are executed (\textit{monotonicity}), because the coverage achieved by previous test cases cannot be eliminated by newly added tests. This means that any coverage function $f$ should satisfy the following property: for all test suites $X$ and $Y$ such that $X \subseteq Y$, it holds that $f(X) \leq f(Y)$. This ensures that even redundant or ineffective tests do not reduce the previously reached coverage value. However, the mathematical formulation of NLC (as defined in Equations~\ref{eq:nlc_l} \& \ref{eq:nlc}) can \textit{decrease} when new tests are executed.
Consider, for instance, a test suite with two test inputs associated with the activation values $+10$ and $-10$ of a given neuron $n$. If we extend this test suite with two more test inputs that trigger the activation values $+5$ and $-5$, the variance $\sigma_n$ of the neuron's output drops from $100$ to $62.5$. However, the new test suite is not less effective than the initial one: it might even be more effective if the inputs with activations $+5$ and $-5$ represent interesting boundary cases.

To address this issue, Yuan et al. proposed updating the coverage value only when it increases, effectively discarding inputs that do not contribute positively to coverage. In our previous example, the two inputs with activations $+5$ and $-5$ would be ignored. While this might look like a pragmatic adjustment to align NLC with the expected behaviour of a coverage criterion, we argue that it is a problematic solution.
This approach can result in \textit{order-dependent} NLC values, where the final coverage depends on the order in which tests are evaluated, even when the same set of tests is considered\footnote{Note that similar issues can arise in conventional coverage criteria if flaky tests are present~\cite{parry2021survey}. However, in such cases, the problem lies not with the coverage criterion itself but rather with the nature of the tests.} (see Section~\ref{sec:exp_order_dep} for our empirical validation). In our running example, if the two tests with activations $+5$ and $-5$ were considered first, they would be retained, as they increase NLC from $0$ (no tests) to $25$. Subsequently, the tests with activations $+10$ and $-10$ would also be retained, further increasing NLC from $25$ to $62.5$. However, if the order is reversed, only the tests with activations $+10$ and $-10$ would be retained, resulting in NLC $= 100$, which differs by $37.5$ from the previous NLC value (the former being hence 37.5\% lower than the latter).

We remark that the discrete rules (e.g., thresholding) introduced in other DNN coverage criteria (e.g., Neuron Coverage) inherently solve this problem, as discrete rules ensure monotonicity by design. \changed{While discrete coverage spaces are common in software testing, we acknowledge that continuous coverage criteria guaranteeing monotonicity and order independence might be possible (e.g., through volume-based measures), though they may present computational challenges and lack intuitive percentage-based interpretations.}

\begin{stripbox}{0.8em}
NLC's mathematical formulation allows its value to decrease when additional tests are executed, violating the monotonicity expected for a coverage criterion. While the authors proposed updating NLC only when it increases, this solution is problematic, as it introduces order-dependent results, further undermining the validity of NLC as a coverage criterion.
\end{stripbox}

The computation of NLC involves summing the absolute values of covariances within the covariance matrix of a layer in a DNN under test. From this, it is evident that NLC is inherently non-negative, with no upper bound. This unbounded nature means that NLC can assume arbitrarily large positive values, which raises further questions about its suitability as a coverage criterion. Ammann \& Offutt~\cite{ammann2017introduction} provided a foundational definition of coverage criterion for conventional software:

\begin{definition}[Coverage Criterion] \label{def:coverage}
\textit{A coverage criterion is a rule or collection of rules that impose specific test requirements on a test set. These rules must describe the test requirements in a complete and unambiguous manner.}
\end{definition}

This definition implies that a coverage criterion serves as a measurable benchmark, enabling developers to set clear testing goals and determine when sufficient testing has been achieved. \changed{It also inherently requires properties such as monotonicity and order independence to reliably track testing progress. For instance, in regression testing, coverage should not decrease when new tests are added to the test suite. Furthermore, NLC's unbounded nature precludes its conversion into a percentage of covered targets, making it incompatible with the traditional notion of a coverage criterion. Consequently, NLC cannot easily be used to define a finite set of test goals and does not naturally lead to a clear stopping criterion for testing.}

Yuan et al. discussed this point in their work, arguing that DNN coverages are not directly comparable to traditional software coverages and suggesting that developers should focus on maximising NLC values through test inputs. However, this does not provide any clear guidance on when to stop testing, %
making NLC impractical for real-world DNN development.\footnote{It is worth noting that even in conventional coverage criteria, achieving full adequacy (e.g., 100\% coverage) is not practical~\cite{marick1999misuse, kurtz2016analyzing}. This is due to factors such as unsatisfiable test goals or goals not worth pursuing. However, in practice, developers can still aim for reasonable coverage levels, focusing on achievable and meaningful targets~\cite{petrovic2018state}.} We agree that NLC would retain utility in guiding automated test generation tools, where the objective is to maximise NLC values (apart from the above mentioned monotonicity problem). However, as Kaufman et al.~\cite{kaufman2022prioritizing} have highlighted -- drawing a distinction between mutation testing and mutation analysis -- this application aligns more closely with research-oriented tasks (akin to mutation analysis, rather than mutation testing~\cite{kaufman2022prioritizing}). Specifically, NLC may be better suited for comparing the effectiveness of different test sets (again, modulo the lack of monotonicity) rather than serving as a practical tool for developers to interpret and improve test coverage. %

The failure of NLC to qualify as an easily interpretable and reliable coverage criterion can be attributed to its design choices. Specifically, NLC avoids discretising the neuron output space (e.g., through thresholding) to satisfy two of its requirements: R1 (precisely describing continuous neuron outputs) and R8 (being hyper\-pa\-ra\-meter-free). While these design decisions confer significant advantages, such as preserving the continuous nature of neuron activations and eliminating the need for manual parameter tuning, they come at the cost of sacrificing NLC's ability to function as a measure of the proportion of test goals that have been achieved.

This investigation raises a broader question: \textit{Can a DNN coverage criterion operate in a continuous way, without hyperparameters and without defining bounds for a discrete set of regions to be covered?} This challenge distinguishes DNN testing from traditional software testing, where coverage criteria (e.g., statement or branch coverage) operate on discrete, well-defined units. In DNNs, however, the absence of such discretisable units complicates the development of coverage criteria that satisfy both R1 and R8.

\begin{stripbox}{0.5em}
While future work might develop monotonic and order-independent variants of NLC (potentially through continuous measures like volume coverage), the current formulation remains limited as a practical test goal for development due to its design choices to satisfy R1 and R8. This underscores the broader challenge of defining DNN coverage criteria in continuous spaces without relying on hyperparameters and discrete regions, while maintaining properties essential for practical use.
\end{stripbox}

\subsection{Does NLC Reflect the Properties of a Covariance Matrix and Layer-Wise Information?}

Beyond our arguments over whether NLC qualifies as a valid coverage criterion, this section critically examines the mathematical design of NLC in its attempt to satisfy R2 (characterising correlations of neurons). We highlight the discrepancies between the properties of a covariance matrix and those of NLC. These discrepancies were conflated in the work by Yuan et al.
We explain why we believe their design choices are problematic, how the properties were misinterpreted (Section~\ref{sec:loss_of_cov_properties}), and propose potential improvements (Section~\ref{sec:new_formulation}). Additionally, we address issues arising from NLC's summation across multiple layers, which we argue leads to unintended and misleading results (Section~\ref{sec:loss_of_layer_info}).

\subsubsection{Loss of Covariance Matrix Properties in NLC}
\label{sec:loss_of_cov_properties}

As outlined in Section~\ref{sec:background_design_nlc}, a covariance matrix encapsulates critical data properties such as divergence, correlation, and shape. However, NLC's formulation (Equation~\ref{eq:nlc_l}) fails to preserve these properties due to its reliance on the L1-norm. Yuan et al. sought to derive a scalar measure representing multiple neurons' multidimensional data, beginning with single-neuron variance (Equation~\ref{eq:single_neuron}). To generalise this to multiple neurons, they needed a measure of the \textit{variance of a covariance matrix}, which describes multi-dimensional data variability. However, NLC simply sums the absolute values of all entries of the covariance matrix, labelling this sum as the layer's NLC. This approach yields a value that is challenging to interpret and lacks meaningful mathematical properties, as it merely combines variances and covariances using summation. Importantly, while a covariance matrix can represent data structure and shape, NLC loses this capability, despite the authors' intent. %

Consider three neurons, $n_1$, $n_2$, and $n_3$, and the following $3\times 3$ covariance matrices:

\[
\Sigma_1 = \begin{bmatrix} 4 & 1 & 0 \\ 1 & 4 & 1 \\ 0 & 1 & 4 \end{bmatrix} \quad \text{and} \quad \Sigma_2 = \begin{bmatrix} 4 & 0 & 2 \\ 0 & 4 & 0 \\ 2 & 0 & 4 \end{bmatrix}
\]

Both matrices share the same L1-norm but represent distinct correlation structures. $\Sigma_1$ exhibits a sequential dependency, where $n_1$ correlates with $n_2$, and $n_2$ correlates with $n_3$, but $n_1$ and $n_3$ are uncorrelated. This structure resembles a chain, where information flows sequentially from one neuron to the next. Its eigenvalues (5.41, 2.59, 4) and eigenvectors reflect this pattern. 
In contrast, $\Sigma_2$ displays a direct correlation between $n_1$ and $n_3$, with $n_2$ being independent of both. Unlike $\Sigma_1$, where correlations are sequential, $\Sigma_2$ shows a pattern where $n_1$ and $n_3$ are directly connected, while $n_2$ remains uncorrelated with either. Its eigenvalues (2, 4, 6) and eigenvectors highlight this pattern. This example illustrates the L1-norm's inability to capture structural nuances, as it cannot distinguish between these two distinct correlation patterns.

Moreover, NLC's handling of the covariance matrix (which is symmetric) results in double-counting of non-diagonal covariances, overemphasising their contribution compared to diagonal variances. This oversight, not addressed by Yuan et al., highlights the need for a revised scoring function that better retains the properties of the covariance matrix, even if some information loss is unavoidable when condensing it into a single scalar value.

\begin{stripbox}{0.8em}
NLC's reliance on the L1-norm fails to preserve important structural and relational properties of the covariance matrix, limiting its interpretability and utility.
\end{stripbox}

\subsubsection{An Alternative Formulation Leveraging Intrinsic Properties of the Covariance Matrix}
\label{sec:new_formulation}

To address the limitations of the L1-norm, we propose a refined scoring function to summarise the covariance matrix into a scalar that aligns more closely with the intrinsic properties of the covariance matrix. Among the various potential formulations, we argue that the \textit{determinant} of the covariance matrix is the most suitable choice. The determinant, defined as:
\begin{equation}
    \text{Det}(\Sigma) = \prod_{i} \lambda_i
\end{equation}
where $\lambda_i$ represents the eigenvalues of $\Sigma$, quantifies the generalised variance of the data. It encapsulates the volume of the data distribution and is intrinsically linked to the Fisher Information matrix, which measures the amount of information an observable random variable carries about an unknown parameter~\cite{lehmann2006theory, van2000asymptotic}. This connection has made the determinant a widely adopted metric in numerous prior studies~\cite{salibian2006principal, hallin2008optimal, dumbgen2005breakdown}.

The determinant offers several key advantages. First, it comprehensively captures both the scale and shape of the data distribution. Second, it is inherently sensitive to correlations, as it depends on the eigenvalues of the covariance matrix. Third, it avoids the double-counting of covariances, eliminating the bias from the L1-norm. However, it is important to note that the determinant is computationally more costly than the L1-norm, particularly for larger matrices. Despite this drawback, its ability to preserve important properties of the covariance matrix makes it a compelling choice.

\changed{We acknowledge, however, that even with our proposed alternative measure using the determinant, NLC would still face the same fundamental issues we have outlined. Nevertheless, compared to NLC, this new formulation more effectively captures the data distribution and the properties of a covariance matrix. Hence, we deem it a promising direction of investigation, although it is not an ultimate solution that improves all of NLC’s deficiencies.}

Alternatively, other scoring functions, such as the \textit{trace} or the \textit{spectral norm} (the largest eigenvalue of the covariance matrix), can be considered. Each of these alternatives comes with distinct trade-offs. The trace of the covariance matrix is computationally efficient and directly measures the total variance. However, it ignores off-diagonal elements (covariances), losing critical information about correlations and failing to capture the shape or orientation of the data distribution. The spectral norm, which reflects the dominant variance direction, provides insight into the scale of the data along its principal axis. Yet, it neglects contributions from other eigenvalues, resulting in a loss of information about the overall shape of the distribution and has a reduced sensitivity to correlations compared to the determinant. While the trace and spectral norm offer computational simplicity, they fail to capture the full structural nuances of the covariance matrix. In contrast, the determinant provides a more comprehensive representation of the data distribution, making it the preferred choice despite its higher computational cost.

\begin{stripbox}{0.8em}
We propose the determinant of the covariance matrix as an improved scoring function over NLC. While computationally intensive, it addresses the limitations of the L1-norm and it outperforms the alternatives in preserving the structural covariance properties.
\end{stripbox}

\subsubsection{Loss of Layer-Wise Information in NLC}
\label{sec:loss_of_layer_info}

As the title of Yuan et al.'s paper suggests, NLC is a \textit{layer-wise} criterion, meaning that it is designed to capture layer-wise information of a DNN. Here, we investigate whether this layer-wise information is effectively preserved in NLC. 

The final NLC aggregates values across all layers of a DNN (see Equation~\ref{eq:nlc}). As a hyperparameter-free metric (satisfying R8), NLC eliminates the need to select a specific layer for coverage calculation. This design aligns with R3, which states that DNN coverage should analyse the distribution of neuron outputs within each layer, as NLC incorporates per-layer neuron distribution. While aggregating values across all layers may seem intuitive, we argue that this approach (summation) overlooks the variability in covariance matrix values between layers. This can result in a single layer disproportionately influencing the overall NLC value, potentially due to differences in the activation function's output ranges or the presence of layers with significantly higher neuron activations. Such dominance can mask the contributions of other layers, leading to a distorted representation. We empirically validate this claim in Section~\ref{sec:exp_sum_layers}, where we show a pronounced imbalance in layer contributions, with one layer dominating the final coverage value and rendering the NLC contributions of other layers negligible. Given this problem, we cannot deem NLC as a truly layer-wise metric.

\begin{stripbox}{0.8em}
Summing NLC across layers runs into the risk of specific layers dominating the final value, masking other layers' contributions.
\end{stripbox}

\section{Critical Review of the  Empirical Study}
\label{sec:critical_review_study}

In addition to proposing a novel formulation of a new DNN coverage criterion, Yuan et al. include experimental results obtained by applying a unique, ad-hoc empirical study design. %
This section outlines their study and provides a critical review, highlighting potential flaws and misleading aspects of their empirical approach.

The empirical study of Yuan et al. consists of two main components: (1) \textit{Assessing Test Suite Quality} and (2) \textit{Guiding Input Mutation in DNN Testing}. The goal of these components is to compare the performance of their NLC criterion to other coverage criteria in two key areas: serving as a better proxy for test suite quality and functioning as a more effective guide for input generation. While the second study, \textit{Guiding Input Mutation in DNN Testing}, is a straightforward and properly-designed contribution, we focus on the first study, \textit{Assessing Test Suite Quality}, as it raises significant concerns. This first study is further divided into two sub-studies: (1) \textit{Diversity of Test Suites} and (2) \textit{Fault-Revealing Capability of Test Suites}. In the following subsections, we will introduce and critically evaluate these two sub-studies.

\subsection{Threats to Validity of Evaluation Based on Test Diversity}
\label{sec:tv_diversity}

\begin{table}[t]
\centering
\caption{Test Suites for the Diversity Study}
\label{tab:div_test_suites}

\scalebox{0.83}{
\begin{tabular}{lll}
\toprule
\textbf{Id} & \textbf{Description} & \textbf{Size} \\ 
\midrule
$\texttt{test}$ & Original test set comprising real-world images. & 10,000 \\ 
\midrule
$\texttt{test}_{\times 1}$ & \makecell[l]{A subset of 100 images from $\texttt{test}$, each perturbed \\ with white noise, subsequently scaled to match \\ the cardinality of $\texttt{test}$.} & $|\texttt{test}|$  \\ 
\midrule
$\texttt{test}_{\times 10}$ & \makecell[l]{A subset of 100 images from $\texttt{test}$, each perturbed \\ with white noise, subsequently scaled to \\ ten times the cardinality of $\texttt{test}$.} & $10 \times |\texttt{test}|$  \\ 
\bottomrule
\addlinespace[5pt]
\multicolumn{3}{l}{\textbf{Ground Truth}$^*$:  $\texttt{test} > \texttt{test}_{\times 10} > \texttt{test}_{\times 1}$} \\

\multicolumn{3}{l}{\small{\makecell[l]{$^*$ This ground truth ordering was established by Yuan et al. based on the \textit{diversity} of \\the test suites, implying a corresponding ordering for \textit{coverage values}.}}}
\end{tabular}
}

\end{table}

To compare coverage criteria, Yuan et al. attempted to establish a ground truth for test suite ordering. They argued that absolute coverage values (e.g., NLC = 0.7 vs. NC = 0.5) cannot be meaningfully compared across different criteria, as each criterion operates on a distinct scale and interpretation of coverage. %
To address this, they defined a relative order of coverage values as the ground truth, asserting that a test suite with higher diversity should yield higher coverage values under any effective criterion. The goal of setting this ground truth was to determine which coverage criterion aligns with this predefined order, which is based on the \textit{diversity} of test suites.

Table~\ref{tab:div_test_suites} shows three test suites they created: the original test set ($\texttt{test}$), the $\times 1$ scheme ($\texttt{test}_{\times 1}$), and the $\times 10$ scheme ($\texttt{test}_{\times 10}$). In the $\times 1$ scheme, 100 images are randomly selected from the test data, and white noise within the range $[-0.1, 0.1]$ is added to generate mutated images, ensuring the total number of these images matches the size of the original test data. The $\times 10$ scheme follows the same procedure but produces ten times the number of mutated images. The authors posit the relative order $\texttt{test} > \texttt{test}_{\times 10} > \texttt{test}_{\times 1}$ as the ground truth, 
meaning that the \textit{coverage} of $\texttt{test}$ should exceed that of $\texttt{test}_{\times 10}$ and $\texttt{test}_{\times 1}$ and \textit{coverage} of $\texttt{test}_{\times 10}$ should exceed that of $\texttt{test}_{\times 1}$. They reason that $\texttt{test}$ contains over 10,000 diverse and meaningful real-world samples, while $\texttt{test}_{\times 1}$ and $\texttt{test}_{\times 10}$ are mostly \textit{dummies} as they are generated with white noise and based on only 100 data samples. This design, however, raises questions about its validity and potential to mislead.

Diversity in a test suite refers to the extent to which tests in a test set can be differentiated from one another. In traditional software testing, this differentiation ensures that the tests cover a broader range of the program under test. However, it is important to note that while diversity can be expected to correlate with higher coverage, this relationship is not absolute. For example, a test suite with high diversity might achieve low coverage if it focuses predominantly on corner cases, leaving many nominal cases untested. 

Diversity can be measured in various ways. For instance, Feldt et al.~\cite{feldt2016test} proposed the Test Set Diameter (TSMd), which calculates the distances between tests using Normalized Compression Distance (NCD)~\cite{cilibrasi2005clustering}. When it comes to DNN testing, we argue that diversity can also be computed in multiple ways. For instance, in an image classification problem, one approach is to calculate the Euclidean distance between test images, which partially aligns with how humans perceive differences between images. 
Another option is to measure the Euclidean distance in some latent space (e.g., an auto-encoder's latent space), where changes in values correspond to semantic changes in the image. Yet another approach is to calculate the distances between neuron outputs when the test images are fed into the DNN. This method aligns with the DNN's internal perspective (i.e., its learned feature space) but may not align with human perception or semantic features of the image. For example, two images that produce significantly different neuron outputs might appear very similar to the human eye (e.g., adversarial examples), and vice versa.

This raises an important question: \textit{How can we say that one test set is more diverse than another?} This is a challenging question to answer, as it depends on the perspective adopted by the user. Turning back to the three test suites with this question in mind, \texttt{test}, $\texttt{test}_{\times 1}$, and $\texttt{test}_{\times 10}$: \textit{Can we definitively determine which suite has higher diversity?} The answer depends entirely on the perspective we adopt. While $\texttt{test}_{\times 1}$ or $\texttt{test}_{\times 10}$ may appear to consist mostly of duplicates derived from 100 images, these perturbed inputs can be perceived as \textit{new and diverse} from the perspective of the DNN. This challenges the assumption that $\texttt{test}$ is inherently more diverse than $\texttt{test}_{\times 1}$ or $\texttt{test}_{\times 10}$. Consequently, the proposed ground truth based on a relative order of diversity may not hold universally and could be misleading. Yuan et al. compare NLC with other methods using this ground truth, which introduces a significant threat to the validity of the comparison. \changed{In support of our argument, we present a counterexample in Section~\ref{sec:counterexample_diversity}, using a diversity measure based on spectral analysis of neuron activations~\cite{humbatova2024spectral} and clustering-based methods~\cite{zohdinasab2024focused}.}

\begin{stripbox}{0.8em}
The proposed ground truth for test suite diversity may not hold, as test suites generated by perturbing a small subset of images can still appear diverse to a DNN. 
\end{stripbox}

\subsection{Threats to Validity of Evaluation Based on Fault Revealing Capability}
\label{sec:tv_fault_revealing}

\begin{table}[t]
\centering
\caption{Test Suites for Fault Revealing Study}
\label{tab:fl_test_suites}

\scalebox{0.88}{
\begin{tabular}{ll}
\toprule
\textbf{Id} & \textbf{Description} \\ 
\midrule
Test & Original test set with 10,000 real-world images. \\ 
\midrule
PGD & \makecell[l]{PGD adversarial examples that deceive the model.} \\
\midrule
CW & \makecell[l]{CW adversarial examples that deceive the model.} \\ 
\midrule
AP & \makecell[l]{Adversarially perturbed inputs created by adding CW or PGD \\ perturbations to 1,000 randomly selected test inputs, \textit{while} \\\textit{preserving their original predictions} (thus not true adversarial \\examples).} \\ 
\bottomrule
\addlinespace[5pt]
\multicolumn{2}{l}{\textbf{Ground Truth}$^*$: PGD $>$ Test, CW $>$ Test, Test $>$ AP $>$ 0} \\

\multicolumn{2}{l}{\small{\makecell[l]{$^*$ This ground truth ordering was established by Yuan et al. based on the \textit{fault-}\\\textit{revealing} \textit{capability} of the test suites, implying a corresponding ordering for \\\textit{coverage values}.}}}
\end{tabular}
}
\end{table}

Faults in DNN testing can take on various meanings. They may include adversarial examples, inputs causing mispredictions or misbehaviours, coding errors introduced by developers, or incorrectly configured hyperparameters of the model~\cite{humbatova2020taxonomy}. Despite this variety, most studies on DNN coverage testing evaluate their criteria based on how effectively they respond to adversarial examples~\cite{ma2018deepgauge, kim2019guiding}. These adversarial examples are generated using white-box attack methods, which leverage gradient information to produce changes that are imperceptible to humans yet capable of altering the model's predictions.\footnote{Note that adversarial examples differ from white noise used in the diversity study in Section~\ref{sec:tv_diversity}. While white noise is random and untargeted, adversarial examples are carefully crafted to mislead the model.} Yuan et al. also treated adversarial examples as faults and evaluated NLC using them. They focused on two well-known adversarial attack methods, as outlined in Table~\ref{tab:fl_test_suites}: Projected Gradient Descent (PGD)~\cite{madry2017towards} and Carlini \& Wagner (CW)~\cite{carlini2017towards}. Using these methods, they generated adversarial examples from either a training set or a test set with a high attack success rate, resulting in two sets named PGD and CW. They established a ground truth stating that PGD $>$ Test and CW $>$ Test, meaning that the \textit{coverage} of PGD and CW should exceed that of the original test set.

However, this formulation of the ground truth is questionable: \textit{Should a test set that primarily covers edge (e.g., adversarial) cases inherently have higher coverage than a test set that focuses on in-distribution data?} We argue that coverage is not synonymous with fault-revealing capability. 
A test set with high fault-revealing capability might actually exhibit lower coverage if it focuses on edge cases rather than broadly covering the input space. In traditional software testing, a single, fault-revealing test has typically lower coverage than a comprehensive test suite that covers all nominal cases, in which no error is triggered.
In fact, focusing on adversarial examples only might leave the DNN exposed to failures associated with boundary or tricky in-distribution inputs that cannot be generated by any adversarial method.
From a functional testing perspective (as compared to security testing), it is equally or even more important to ensure that the DNN behaves correctly in the vast majority of in-distribution and boundary cases, rather than considering just adversary manipulations. 
This reasoning can also be extended to the diversity study discussed in Section~\ref{sec:tv_diversity}: 
a highly diverse test set might have relatively low coverage, if it focuses only on corner cases, leaving most nominal cases untested.

\begin{stripbox}{0.8em}
Coverage $\neq$ fault-revealing capability: a test set consisting only of edge-case tests may have high fault detection, but low coverage. A good test set should include both nominal cases, boundary cases, corner cases and adversarial cases, not just some of these categories.
\end{stripbox}

In their study, Yuan et al. found that many existing coverage criteria satisfy their ordering of PGD $>$ Test and CW $>$ Test. To further investigate this aspect, they introduced a new test set called AP (see Table~\ref{tab:fl_test_suites}), created by adding adversarial perturbations (CW or PGD) to 1,000 randomly selected inputs from the test set (which contains 10,000 inputs) without changing their predictions. Thus, AP is not adversarial. They established a ground truth ordering of Test $>$ AP $>$ 0, meaning that \textit{coverage} of Test should exceed that of AP. They reason that Test is ten times larger than AP and AP does not cause mispredictions, and showed that only NLC satisfied this ordering, while other criteria failed, claiming those criteria did not accurately reflect fault-revealing capability but were instead sensitive to the dissimilarity induced by adversarial perturbations, even though Test is larger than AP.

However, this ground truth ordering is also problematic. First, as previously argued, coverage is not equivalent to fault-revealing capability. Second, even though AP does not cause mispredictions, it does push the inputs closer to the decision boundaries of the DNN through adversarial perturbations, bringing them near to the edge cases. Therefore, we argue that such adversarial perturbations should be viewed positively, as they contribute to DNN testing in areas near or inside the in-distribution data, where regressions may occur when the DNN evolves (e.g., through successive training on new data). \changed{To support our argument, we present a counterexample in Section~\ref{sec:counterexample_fault}, where we use DNN mutants~\cite{humbatova2021deepcrime} as proxies for real faults and compare the fault-detection effectiveness of Test and AP based on their mutant-killing capabilities.}

\begin{stripbox}{0.8em}
Yuan et al. introduced AP, a test set with adversarial perturbations that do not cause mispredictions, and claim that this is (by construction) a weak test set. We instead argue that AP perturbations are not necessarily useless, as they bring the inputs closer to the decision boundaries, where regressions may occur in the future.
\end{stripbox}

\section{Empirical Validation of Our Arguments}
\label{sec:empirical_validation}

Our primary goal in this section is to perform a targeted study on the theoretical foundations and empirical evaluation of NLC, as outlined in Section~\ref{sec:critical_review_nlc}, rather than to replicate all extensive experiments conducted by Yuan et al.

\subsection{Experimental Setup}
We utilise the same set of models and datasets for classification as Yuan et al.: two datasets, CIFAR-10 and ImageNet, and three DNN architectures with their corresponding pre-trained weights, VGG16\_BN, ResNet50, and MobileNet\_V2, all implemented in PyTorch. For further details on these models and datasets, consult Yuan et al.~\cite{nlcpaper}.

\begin{table}[h!]
\centering
\caption{Instability of NLC under Data Shuffling} \label{tab:variance_summary}
\scalebox{0.82}{
\begin{tabular}{c|c|c|c|c|c}
    \toprule
    \textbf{Dataset} & \textbf{Shuffled?} & \textbf{Model} & \textbf{Std} & \textbf{SEM} & \textbf{Max \% Drop} \\
    \midrule
    \multirow{6}{*}{CIFAR-10} & No & ResNet50 & 0.0 & -- & 0.0 \\
     & No & VGG16\_BN & 0.0 & -- & 0.0 \\
     & No & MobileNet\_V2 & 0.0 & -- & 0.0 \\
     \cmidrule{2-6}
     & Yes & ResNet50 & \textbf{0.0068} & \textbf{0.0015} & \textbf{1.58\%} \\
     & Yes & VGG16\_BN & \textbf{0.0304} & \textbf{0.0068} & \textbf{4.22\%} \\
     & Yes & MobileNet\_V2 & \textbf{0.0044} & \textbf{0.0010} & \textbf{0.84\%} \\
    \midrule
    \multirow{6}{*}{ImageNet} & No & ResNet50 & 0.0 & -- & 0.0 \\
     & No & VGG16\_BN & 0.0 & -- & 0.0 \\
     & No & MobileNet\_V2 & 0.0 & -- & 0.0 \\
     \cmidrule{2-6}
     & Yes & ResNet50 & \textbf{0.0681} & \textbf{0.0152} & \textbf{7.46\%} \\
     & Yes & VGG16\_BN & \textbf{0.1266} & \textbf{0.0283} & \textbf{10.42\%} \\
     & Yes & MobileNet\_V2 & \textbf{0.1284} & \textbf{0.0287} & \textbf{8.74\%} \\
    \bottomrule
\end{tabular}
}
\end{table}

\subsection{Order-Dependency of NLC}
\label{sec:exp_order_dep}

As highlighted in Section~\ref{sec:is_nlc_valid}, we hypothesise that NLC can be sensitive to the order in which test data is fed into the model. To test this, we first compute NLC by feeding the test set into the model without any shuffling, establishing a baseline for comparison. Next, we shuffle the test set, randomising the order of data points, and compute NLC again. This process is repeated 20 times to assess the variability in NLC values. Finally, we measure the standard deviation of the NLC values across the 20 runs to quantify the impact of data order on NLC.

The results are summarised in Table~\ref{tab:variance_summary}. The column `Shuffled?' indicates whether the test set was shuffled, while `Std' reports the standard deviation of the NLC values across the 20 runs. \changed{The `SEM' column shows the standard error of the mean, quantifying the precision of the mean estimate.} The `Max \% Drop' column complements the standard deviation by quantifying the worst-case reduction in NLC values due to input order, calculated as $\frac{max - min}{max} \times 100$, where $max$ and $min$ are the maximum and minimum NLC values observed across the 20 runs, respectively.

When the test set is not shuffled, the NLC values remain consistent, with a standard deviation of 0.0 across all models and datasets. In contrast, shuffling the test set introduces variability in the NLC values, as evidenced by non-zero standard deviations. For the CIFAR-10 dataset, the max \% drop ranges from 0.84\% to 4.22\%, while for the more complex ImageNet dataset, it increases significantly, ranging from 7.46\% to 10.42\%.

\changed{To further validate the reliability of using 20 runs, we computed the Standard Error of the Mean (SEM) and relative SEM (i.e., SEM divided by the sample mean) for the NLC values. SEM measures the precision of the estimated mean. A low relative SEM (typically below 5 to 10\%) indicates that the mean is a stable estimate. For CIFAR-10, the SEM (and relative SEM) were 0.0015 (0.08\%), 0.0068 (0.24\%), and 0.0010 (0.06\%) for ResNet50, VGG16\_BN, and MobileNet\_V2, respectively. Similar trends were observed for ImageNet. These results suggest that 20 repetitions provide a sufficiently accurate and stable estimate of the metrics of interest.}

These findings confirm that NLC is indeed sensitive to the order of the inputs. We attribute this behaviour to the design of NLC, which updates its coverage value only when new test inputs increase the coverage, discarding those that do not. This design choice, while intended to enforce monotonicity, inadvertently introduces non-determinism into the coverage metric. We argue that such non-determinism is problematic, as it undermines the reliability of NLC as a consistent measure of test coverage, even in the absence of flaky tests.

\subsection{Layer Aggregation of NLC}
\label{sec:exp_sum_layers}

\begin{figure}[t]
    \centering
    \includegraphics[width=\linewidth]{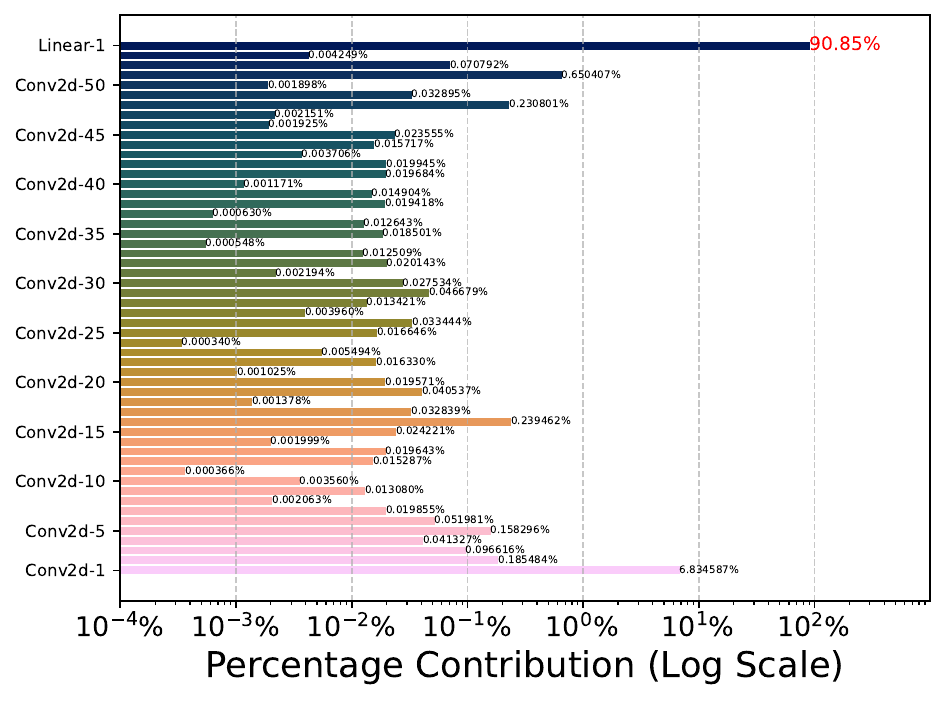}
    \caption{Visualisation of layer contributions in a log scale of ResNet50 on ImageNet}
    \label{fig:layers_sum}
\end{figure}

As discussed in Section~\ref{sec:loss_of_layer_info}, to empirically validate our hypothesis that NLC loses layer-wise information due to its aggregation mechanism, we analyse the contributions of individual layers to the final NLC value. %

Figure~\ref{fig:layers_sum} illustrates the layer-wise contributions to the final NLC value for the ImageNet dataset with ResNet50.\footnote{For brevity, we report full results in our artefact~\cite{misc:artifact} as they exhibit similar trends.} The x-axis uses a logarithmic scale, and each bar is annotated with the percentage contribution of the corresponding layer. The y-axis represents the layers, ordered from the deepest (top) to the shallowest (bottom).
The results reveal a significant imbalance in layer contributions, with the last layer consistently dominating the final NLC value. Specifically, the last layer contributes over 89\% of the total NLC value in all cases, rendering the contributions of the remaining layers negligible. This pattern is consistent across all models and datasets.

These findings substantiate our argument that NLC fails to preserve layer-wise information effectively. By aggregating values across all layers, NLC obscures the variability in covariance matrix values between layers, resulting in a distorted and non-representative measure of coverage. This undermines the claim of NLC being a layer-wise metric, as it cannot reliably capture the distinct behaviours and contributions of individual layers within a DNN. Consequently, we conclude that while NLC simplifies coverage calculation by eliminating the need for layer selection, it sacrifices critical layer-specific insights, limiting its effectiveness as a comprehensive coverage criterion.

One might argue that this imbalance is not problematic, as it naturally reflects the different variances in neuron outputs across layers. Additionally, the unintentional focus on the deepest layer could be seen as justified, given that deeper layers are known to encode more abstract and meaningful features~\cite{bengio2013better}. This reasoning aligns with previous work that, when requiring the choice of a layer, often defaults to the deepest layer as the target~\cite{kim2019guiding}. However, we contend that Yuan et al. did not aim for this outcome; rather, their goal was to incorporate all layer-wise information into a single coverage metric. Unfortunately, this incorporation did not achieve the desired effect. We further explore potential resolutions to this issue in Section~\ref{sec:takeaways}.

\changed{
\subsection{Counterexamples to the Proposed Ground Truth Ordering}
\label{sec:counterexample}

While our critique focuses on logical inconsistencies in how Yuan et al.\ define ground truth ordering based on test suite diversity and fault-revealing capability, to further empirically validate our claims, we conduct three preliminary experiments. Our goal is to provide counterexamples to their proposed ground truth ordering.

\subsubsection{Fault-Revealing Capability (Test $>$ AP)}
\label{sec:counterexample_fault}

To illustrate a case that violates the assumed ordering based on fault-revealing capability (i.e., Test $>$ AP), we employed DNN mutants generated by DeepCrime~\cite{humbatova2021deepcrime} as proxies for real faults. Using the CIFAR-10 dataset, we generated a total of 101 mutants and created an adversarially perturbed set, AP, by applying adversarial perturbations to the original test set (Test) while preserving the original predictions. Our mutation analysis revealed that Test killed 44 mutants, whereas AP killed 46 mutants, directly contradicting the proposed ordering. This counterexample supports our claim that adversarial perturbations push inputs closer to the decision boundaries even when prediction-preserving, thereby enhancing their potential to uncover faults.

\subsubsection{Test Diversity (\texttt{test} $>$ \texttt{test$_{\times1}$})}
\label{sec:counterexample_diversity}
Next, we address the assumed ground truth ordering based on test diversity (i.e., \texttt{test} $>$ \texttt{test$_{\times1}$}). Although Yuan et al. did not formally define diversity, they constructed \texttt{test$_{\times1}$} by applying white noise to 100 randomly sampled test inputs, scaling it to match the size of the original test set, and treating this dummy set as strictly less diverse. %
To challenge this, we used a spectral analysis to quantify diversity, which captures DNN behaviour through histograms of neuron activation values~\cite{humbatova2024spectral}. On CIFAR-10, we measured the Jensen-Shannon (JS) divergence between averaged activation histograms of different original test sets and dummy sets, repeated over 20 independent model re-trainings.  %
To get the possibility of choosing 100 diverse samples, we apply the K-Means clustering algorithm to the full test set's activation spectrums with K set to 100. Then, for each of 100 clusters, we choose 1 data point closest to its centroid. The resulting subset of 100 samples exhibited a JS divergence of $2.18 \times 10^{-5}$ from the full set.
A dummy set built from these `centroid' samples showed a JS divergence of $7.95 \times 10^{-5}$, indicating it retained most of the original diversity. In contrast, dummy sets based on 100 samples from a single cluster yielded a much higher divergence ($1.14 \times 10^{-2}$). Random sampling also retained strong diversity, averaging $5.87 \times 10^{-5}$ in JS divergence. These results demonstrate that, depending on the selection strategy, dummy sets do not have to be less diverse.

We further evaluated test diversity using the clustering-based strategy proposed by Zohdinasab et al.~\cite{zohdinasab2024focused}. Their approach projects inputs into a lower-dimensional latent space, applies clustering, and measures how well a test set covers the resulting clusters. Applying this method to the CIFAR-10 test set yielded an optimal cluster count of 40. Selecting one sample per cluster led to a 40-input subset that effectively represented the diversity of the full test set under this metric. This demonstrates that even small, carefully constructed subsets can adequately reflect global diversity.
}

\section{Discussion and Implications}
\label{sec:discussion}

This section discusses our findings, revisits DNN coverage criteria, explores their practical use, examines DNN faults, highlights comparison challenges, and summarises key insights from our review.

\subsection{Revisiting the Analysis of Prior DNN Coverage Criteria}
\label{sec:discussion_other_criteria}

Based on Yuan et al.'s analysis of existing DNN coverage criteria, no criteria other than NLC satisfy all eight design requirements, although NLC ultimately fails to qualify as a coverage criterion according to our analysis. Upon examining other criteria, we found that none have similarly fallen short in the same manner as NLC. A related concept to NLC would be  Surprise Adequacy (SA)~\cite{kim2019guiding}, which, like NLC, operates with unbounded values and correspondingly is not classified as a coverage criterion. To transform SA into a coverage criterion called Surprise Coverage (SC), Kim et al.~\cite{kim2019guiding} introduced lower and upper bounds to partition the continuous space into discrete buckets, which then serve as test requirements to be covered. Regarding experimental designs to evaluate DNN coverage criteria, we found that most studies have focused on their ability to expose faults or on their utility in guiding test generation tools. The evaluation approach proposed by Yuan et al. with an ad-hoc ground truth ordering is unique, but also problematic, as evidenced by our analysis.

\subsection{Practical Applicability of DNN Coverage Criteria}
\label{sec:discussion_practical_applicability}

We previously mentioned that achieving full adequacy (e.g., 100\% coverage) is impractical, even in traditional software testing due to factors such as unsatisfiable or low-priority test goals~\cite{marick1999misuse, kurtz2016analyzing}. In practice, developers must prioritise and refine test objectives to focus on meaningful and achievable coverage targets~\cite{petrovic2018state}. This is facilitated by various heuristics and reduction techniques, such as mutant reduction or selectively testing critical functions and methods. When it comes to DNN coverage testing, a similar strategy can be applied by focusing on specific neurons or layers, responsible for faulty behaviours~\cite{sohn2023arachne, sun2022causality}. A key research challenge lies in balancing efficiency and comprehensiveness: leveraging targeted testing techniques may expedite the process, but it is crucial to assess the trade-offs involved in potentially sacrificing full DNN coverage. Investigating these trade-offs could provide valuable insights into optimising DNN coverage testing. Moreover, the notion of coverage regions/targets is very natural and interpretable for developers. It also automatically ensures monotonicity and order independence. Hence, the introduction of a continuous criterion for DNN coverage should not sacrifice such practically important properties.

\subsection{A Broader Perspective on DNN Faults}
\label{sec:discussion_dnn_faults}

To the best of our knowledge, existing DNN coverage testing works have primarily focused on adversarial examples, with little to no evaluation of other types of DNN faults, such as inputs that are not adversarial but still cause mispredictions. We argue that this focus is largely due to adversarial examples being introduced earlier than other fault types and their widespread recognition as significant threats to DNNs. However, as reported by Humbatova et al.~\cite{humbatova2020taxonomy}, developers perceive DNN faults more broadly, including mistakes in PyTorch model code or improper model hyperparameters as faults. DNN mutation testing techniques have explored this broader spectrum of faults~\cite{humbatova2021deepcrime}, demonstrating their effectiveness in testing DNN models by artificially seeding faults that simulate common developer mistakes. DNN coverage criteria could be evaluated directly for their ability to expose such artificially injected faults, providing a more comprehensive assessment of their effectiveness.

\subsection{Difficulties in Comparing Coverage Criteria}
Comparing coverage criteria has been a challenge, both in traditional software testing and, more recently, in DNN testing. As noted by Ammann \& Offutt~\cite{ammann2017introduction}, one common theoretical approach to comparing criteria is through subsumption relationships. Specifically, criterion $C_1$ is said to subsume $C_2$ if every test set that satisfies $C_1$ also satisfies $C_2$. While this provides a foundational framework for comparison, it is not without limitations. For example, if $C_1$ contains infeasible test requirements, conditions that cannot be satisfied by any test case, a test suite meeting $C_1$ might inadvertently skip satisfiable requirements of $C_2$. Ammann \& Offutt argue that, in practice, infeasible requirements are rare, and when they do occur, the corresponding requirements in $C_2$ are often infeasible as well. Although subsumption offers a useful theoretical framework for comparison, it is in practice quite limited. In many cases, the two criteria $C_1$ and $C_2$ will be incomparable, meaning that neither $C_1$ subsumes $C_2$ nor vice versa. For instance, in traditional software testing, coverage and mutation criteria are theoretically incomparable (unless all coverage targets are assumed to be mutated), and the same holds for branch vs. condition coverage (because flipping all conditions does not necessarily flip all decisions). We expect that most DNN coverage criteria cannot be compared for subsumption theoretically and that empirical testing will often reveal them to be incomparable (i.e., neither subsumes the other).

Moreover, while intuition might suggest that a subsuming criterion ($C_1$) should detect more faults than the subsumed criterion ($C_2$), there is no theoretical basis to guarantee this, and empirical studies have yet to provide conclusive evidence. Therefore, we believe the only viable alternative for comparing DNN coverage criteria is their evaluation on a variety of diverse fault types, ideally including real faults that developers encountered when training and evolving DNNs~\cite{humbatova2020taxonomy}. A comprehensive benchmark of real DNN faults is still missing~\cite{jahangirova2024realfaultsdeeplearning}, but it would be extremely beneficial to the field. In the absence of such a benchmark, artificial faults injected by DNN mutation tools remain a viable alternative~\cite{humbatova2021deepcrime}.

\subsection{Key Takeaways from Our Critical Review}
\label{sec:takeaways}

Here, we summarise our review of Yuan et al., aiming to contribute to the development of improved DNN coverage criteria:

\noindent \textbf{\textit{Definition of Coverage Criterion:}} When designing a new DNN coverage criterion, some important foundational principles behind coverage criteria in traditional software testing should be preserved, such as monotonicity, order independence and its usage as a stopping criterion. If a proposed criterion fails to meet these key requirements, it becomes difficult for developers to understand what it measures.
While the eight requirements suggested by Yuan et al. provide valuable insights, we emphasise the importance of maintaining legitimacy as a coverage criterion while striving to satisfy them.

\noindent \textbf{\textit{Layer-Specific Coverage Reporting:}} As highlighted in Sections~\ref{sec:loss_of_layer_info} and ~\ref{sec:exp_sum_layers}, we observed that aggregating NLC coverage across all layers results in a loss of layer-wise information. A simple yet effective solution is to report coverage for each layer individually rather than collapsing it into a single metric. This approach aligns with traditional software coverage practices, where reports are generated at different granularities (e.g., file-level, class-level, function-level) to help developers identify and address low-coverage areas. 
Developers would still remain free to focus on one or more specific layers, which are expected to perform semantically meaningful computations and ignore the coverage measures for the others. However, a potential drawback of this multi-dimensional coverage reporting is that it establishes only a partial order relation, which can render certain pairs of test sets incomparable. For instance, given two test sets $T_1$ and $T_2$, $T_1$ might achieve high coverage on layer $l_1$ but low coverage on layer $l_2$, while the opposite could be true for $T_2$. If a single scalar measure is desirable, our findings suggest that simply summing the coverage values is inadequate unless appropriate normalisation measures are applied to ensure meaningful aggregation.

\noindent \textbf{\textit{Pitfalls in Establishing Ground Truth:}} We caution against setting a predefined ground truth order for test sets when evaluating DNN coverage criteria, as it can be misleading and compromise the validity of the evaluation. 
The definition of what constitutes a strong vs. a weak test set is inherently challenging in general.
However, we propose that an order can be established for monotonic sequences of test sets, provided that the corresponding coverage criteria are also monotonic.
Specifically, a sequence of test sets can be carefully constructed such that each subsequent set is a superset of the previous one.
This allows for the quantification of a coverage criterion's  \textit{sensitivity} to the transition from a weaker to a stronger test set, which can then serve as a basis for comparison~\cite{humbatova2021deepcrime}.

Ultimately, the primary goal of a coverage criterion is to ensure that the test set is sufficiently robust to detect all or most faults that might otherwise emerge in real-world scenarios. Hence, we recommend empirically assessing and comparing how different coverage criteria respond to test sets with different fault-revealing capabilities, considering real or artificially injected faults. Another viable approach is to integrate a coverage criterion into an automated test generator and evaluate its capability to drive the test generator toward the generation of useful, fault-revealing tests. This method allows for the comparison of different coverage criteria based on their capability to guide the test generator toward exploring critical regions of the input space.

\section{Related Work}
\label{sec:related_work}

Previous research has critically examined DNN coverage criteria, revealing important limitations. Recent studies have shown that higher coverage metrics do not always correlate with improved model quality~\cite{yan2020correlations, yang2022revisiting} and are often less effective than adversarial methods~\cite{li2019structural}. Harel et al.~\cite{harel2020neuron} empirically demonstrated that increasing NC can sometimes be counterproductive, as it may generate less natural inputs and even negatively correlate with fault detection. Yuan et al.~\cite{nlcpaper}, our reference work, also revisited existing DNN coverage criteria, identified their shortcomings, and proposed eight design requirements for more effective criteria. Building on Yuan et al.'s contributions, our work critically evaluates their proposed criteria and refines their guidelines. Our goal is to pave the way for the development of more robust and effective DNN coverage criteria.

\section{Conclusion}
\label{sec:conclusion}

We conducted a critical review of the DNN coverage criterion proposed by Yuan et al.~\cite{nlcpaper}, questioning some of their theoretical and empirical assumptions. Our analysis revealed that NLC fails to satisfy some of the core properties of a coverage criterion, such as monotonicity, order independence, and usability as a stopping criterion, despite the authors’ efforts to meet eight carefully defined design requirements. Additionally, we identified issues in the empirical study design, particularly the reliance on an unreliable ground truth for evaluating the test suites. Our critique guides for improving future DNN coverage metrics, emphasising the need for evaluation methods that align with the practical value a coverage criterion is expected to deliver to developers.

\section*{Data Availability}
\label{sec:data_availability}
The code and data for our empirical validation are publicly available~\cite{misc:artifact}.

\begin{acks}
We would like to express our sincere gratitude to Robert Feldt for his valuable feedback on our ideas and his thoughtful encouragement.
Shin Yoo has been supported by the National Research Foundation of Korea grant (RS-2023-00208998) and the Engineering Research Center Program (RS-2021-NR060080), funded by Korean Government (MSIT). This research was supported by Basic Science Research Program through the NRF funded by the Ministry of Education (RS-2024-00411975).
\end{acks}

\bibliographystyle{ACM-Reference-Format}
\balance
\bibliography{bib}

\end{document}